\documentclass[preprint,3p]{elsarticle}

\usepackage{amssymb}

\usepackage{graphicx}
\usepackage{subfigure}
\usepackage{amsmath}
\usepackage{multirow}

\journal{arXiv.org}

\begin{document}

\begin{frontmatter}



\title{Bypass rewiring and extreme robustness of Eulerian networks}


\author[address1]{Junsang Park\corref{mycorrespondingauthor}}
\cortext[mycorrespondingauthor]{Corresponding author.}
\ead{junsp85@kaist.ac.kr}
\author[address1]{Seungwon Shin}
\author[address2]{Sang Geun Hahn}
\address[address1]{School of Electrical Engineering, Korea Advanced Institute of Science and Technology, 291 Daehak-ro, Yuseong-gu, Daejeon 34141, Republic of Korea}
\address[address2]{Department of Mathematical Sciences, Korea Advanced Institute of Science and Technology, 291 Daehak-ro, Yuseong-gu, Daejeon 34141, Republic of Korea}

\begin{abstract}
Bypass rewiring improves connectivity and robustness of networks against removal of nodes including failures and attacks.
A concept of bypass rewiring on directed networks is proposed, and random bypass rewiring on infinite directed random networks is analytically and numerically investigated with simulations.
As a result, random bypass rewiring makes infinite directed (undirected) random networks extremely robust for arbitrary occupation probabilities if and only if in-degree of every node except a fixed number of nodes is equal to the out-degree (every node except a finite number of nodes has even degree); bypass rewiring makes the percolation threshold $0$.
Consequently, a finite network has a strongly connected spanning subnetwork which has an Eulerian path or cycle if and only if there exists an way of bypass rewiring to make the finite network extremely robust for every combination of removed nodes; Eulerian networks are extremely robust for every combination of removed nodes.
\end{abstract}


\end{frontmatter}


\section{Introduction}
Many systems in real-world (the Internet, electric power grids, and others) can be represented by complex networks with many nodes (vertices) and links (edges) between nodes \cite{cohen00, cohen01, albert04}.
Complex networks are relatively robust to failures or errors (random removal of nodes) but fragile and vulnerable to intended attacks (targeted removal of nodes in decreasing order of degree from the highest degree); a network is fragmented into smaller components when nodes are deleted \cite{albert00, callaway00, cohen00, cohen01, albert02, dorogovtsev02, schwartz02, newman03, boccaletti06, dorogovtsev08, newman}.
Even though there are various mitigation methods attempted to improve robustness of networks, they have technical, economic, or geographical problems and limitations when applied to real-world systems \cite{beygelzimer05, xiao10, schneider11, quattrociocchi14, shang15}.
From the practical point of view, bypass rewiring is a technicaly, economically, and geographically realistic mitigation method against removal of nodes including failures and attacks because bypass-rewiring the links of a removed node under failures or attacks is easy and simple work; an engineer or equipment can easily and simply rewire cables (links) of a router (node) and repeat the signals directly when the router does not work under failures or attack or is under repair \cite{park16}.

In this paper, we propose a concept of bypass rewiring on directed networks and give generalized results on not only infinite networks but also real-world networks with a finite number of nodes.
In Section \ref{sec:concept}, a concept of bypass rewired is proposed.
In Section \ref{sec:random}, we derive analytical and simulation results of random bypass rewiring on infinite directed random networks.
In Section \ref{sec:discussion}, the results in Section \ref{sec:random} and \cite{park16} with a real-world example, the Internet topology, are discussed and generalized results on infinite random networks and finite networks are derived from the discussion.
In Section \ref{sec:conclusion}, we summarize the paper and comment on further work.

\section{A concept of bypass rewiring on directed networks} \label{sec:concept}
\begin{figure}
\center
\subfigure[]{
	\includegraphics[width=1in]{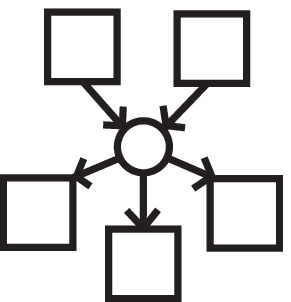}
	\label{fig:before_removal_directed}
}
\subfigure[]{
	\includegraphics[width=1in]{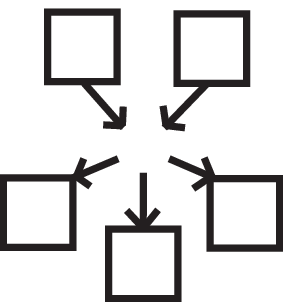}
	\label{fig:without_bypass_directed}
}
\subfigure[]{
	\includegraphics[width=1in]{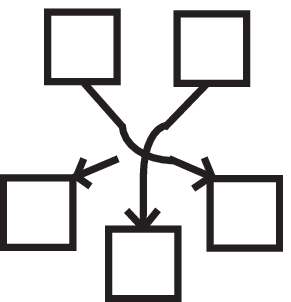}
	\label{fig:with_bypass_directed}
}
\caption{\label{fig:bypass_rewiring_directed} (a)~Before removal of the node, one node (circle) and five components (square) are connected. (b)~After removal of the node, the network fragments into five smaller components without bypass rewiring. (c)~After removal of the node, the network fragments into two larger connected components and one smaller out-component with bypass rewiring. }
\end{figure}

We propose a concept of bypass rewiring on directed networks.
A node in Fig.~\ref{fig:before_removal_directed} is removed by failures or attacks and turns into the removed nodes in Fig.~\ref{fig:without_bypass_directed}.
Bypass rewiring on a directed network is to directly connect each pair of in-links and out-links of the removed node like Fig.~\ref{fig:with_bypass_directed}.
Each pair of in-links and out-links for rewiring can be chosen in various ways including random bypass rewiring by which each pair of in-links and out-links of the removed nodes are randomly chosen.
If in-degree $k_\text{in}$ is larger (smaller) than out-degree $k_\text{out}$ of the removed node, $k_\text{in} - k_\text{out}$ in-links ($k_\text{out} - k_\text{in}$ out-links) remain open.

\section{Random bypass rewiring on infinite directed random networks} \label{sec:random}
\subsection{Analytical results}
Using double generating functions based on the generating function formalism introduced in \cite{callaway00, newman01, schwartz02, newman, wilf}, we define
\begin{flalign}
\qquad & G_{0,0}(x,y) = \sum_{j=0}^{\infty} {\sum_{k=0}^{\infty} {p_{j,k} x^j y^k} }\text{,} & \label{eq:G_0_0}
\end{flalign}
\begin{subequations}
\begin{flalign}
\qquad & H_{\text{in},1}(x) = \sum_{k=0}^{\infty} {h_{\text{in},k} x^k } \text{,} & \label{eq:H_in_1} \\
\qquad & H_{\text{out},1}(x) = \sum_{k=0}^{\infty} {h_{\text{out},k} x^k } \text{,} & \label{eq:H_out_1}
\end{flalign}
\end{subequations}
for
\begin{flalign}
\qquad & \sum_{j=0}^{\infty}{ \sum_{k=0}^{\infty} { j p_{j,k}  }  } = \sum_{j=0}^{\infty} {jp_{\text{in},j}} = \langle j \rangle = \sum_{j=0}^{\infty}{ \sum_{k=0}^{\infty} { k p_{j,k}  }  } = \sum_{k=0}^{\infty} {kp_{\text{out},k}} = \langle k \rangle \text{,} & \label{eq:j_k}
\end{flalign}
where $p_{j,k}$ is the probability that a randomly chosen node has in-degree $j$ and out-degree $k$, and $h_{\text{in},k}$ ($h_{\text{out},k}$) is the probability that a randomly chosen link originates from (leads to) a small in-component (out-component) which has $k$ nodes.
Eq.~(\ref{eq:j_k}) is naturally assumed since average in-degree $\langle j \rangle$ and average out-degree $\langle k \rangle$ are equal on directed networks.
Since nodes of the giant strongly connected component do not belong to any small in- and out-component which has a fixed number of nodes on an infinite directed random networks, the probability that a randomly chosen node belongs to the giant strongly connected component is
\begin{flalign}
\qquad & S_\text{s} = \sum_{j=0}^{\infty} { \sum_{k=0}^{\infty} { p_{j,k} \phi_{j,k} (1 - u_\text{in}^j - u_\text{out}^k + u_\text{in}^j u_\text{out}^k)    }    } \text{,} & \label{eq:S_s}
\end{flalign}
for
\begin{subequations}
\begin{flalign}
\qquad & H_{\text{in},1}(x) = \sum_{j=0}^{\infty} \sum_{k=0}^{\infty} \frac{ (k+1) p_{j,k+1}  }   {\langle k \rangle} \{ 1 - \phi_{j,k+1} + \phi_{j,k+1} [H_{\text{in},1}(x)]^j \} \text{,} \label{eq:H_in_1_H_in_1} & \\
\qquad & H_{\text{out},1}(x) = \sum_{j=0}^{\infty} \sum_{k=0}^{\infty} \frac{ (j+1) p_{j+1,k} } {\langle j \rangle} \{ 1 - \phi_{j+1,k} + \phi_{j+1,k} [H_{\text{out},1}(x)]^k \} \text{.} & \label{eq:H_out_1_H_out_1}
\end{flalign}
\end{subequations}
$u_\text{in}$ ($u_\text{out}$) is the average probability that there exists no path from (to) the giant strongly connected component to (from) the node from (to) which a randomly chosen link originates (leads) where $u_\text{in}$ and $u_\text{out}$ are the smallest non-negative real solutions of
\begin{subequations}
\begin{flalign}
\qquad & u_\text{in} = H_{\text{in},1}(1) = f_1(u_\text{in}) = \sum_{j=0}^{\infty} \sum_{k=0}^{\infty} \frac{ k p_{j,k} (1 - \phi_{j,k} + \phi_{j,k} u_\text{in}^j) }  {\langle k \rangle} \text{,} \label{eq:u_in} & \\
\qquad & u_\text{out} = H_{\text{out},1}(1) = f_2(u_\text{out}) = \sum_{j=0}^{\infty} \sum_{k=0}^{\infty} \frac{  j p_{j,k} (1 - \phi_{j,k} + \phi_{j,k} u_\text{out}^k)   }  {\langle j \rangle} \text{,} & \label{eq:u_out}
\end{flalign}
\end{subequations}
respectively, and $\phi_{j,k}$ is the occupation probability that a randomly chosen node with in-degree $j$ and out-degree $k$ is not removed; the average occupation probability is defined as
\begin{flalign}
\qquad & \phi = \sum_{j=0}^{\infty}{ \sum_{k=0}^{\infty} { p_{j,k} \phi_{j,k}   }    } \text{.} & \label{eq:phi_p_j_k}
\end{flalign}

\begin{figure}[]
\centering
\subfigure[] {
\includegraphics[width=4in]{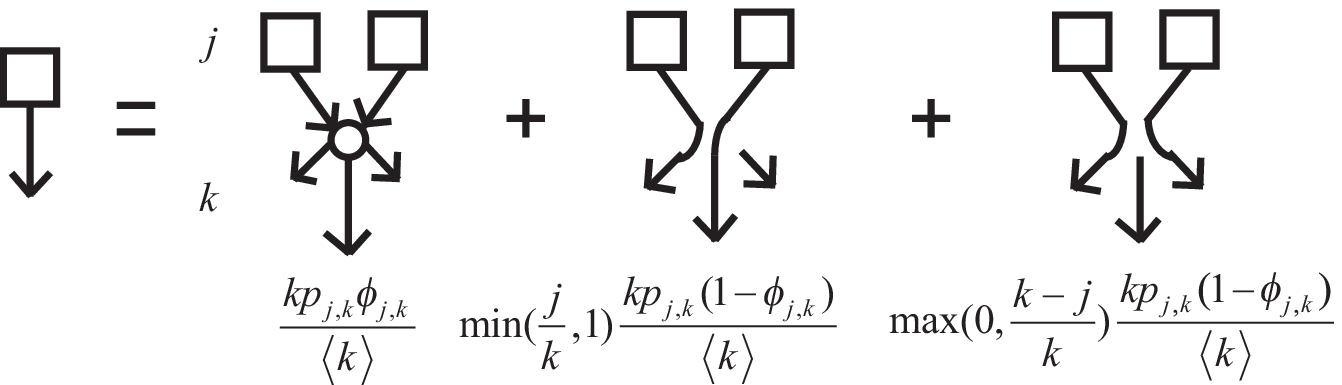}
\label{eq:gen_bypass_in}
}
\subfigure[] {
\includegraphics[width=4in]{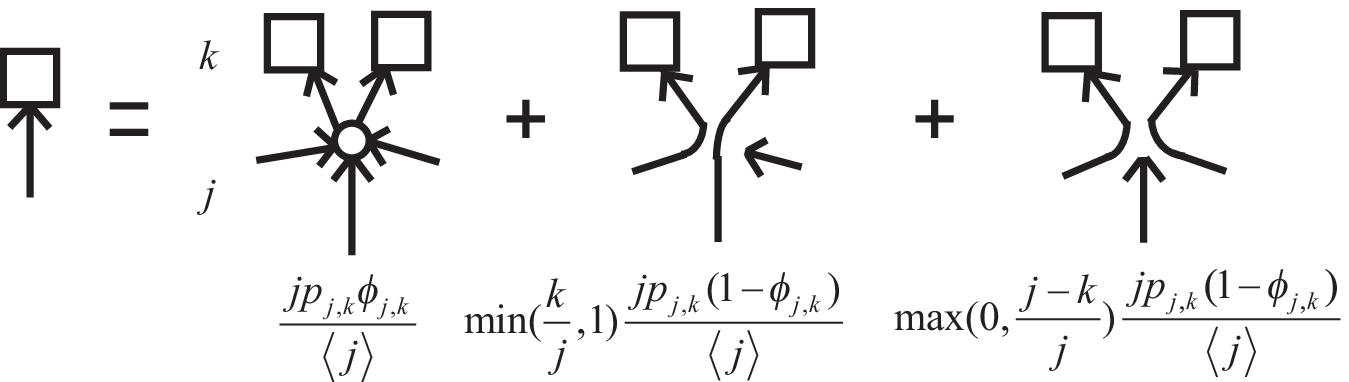}
\label{eq:gen_bypass_out}
}
\caption{\label{fig:gen_bypass_in_out} (a)~A schematic diagram to calculate the probability that a randomly chosen in-link originates from a small in-component [square] with random bypass rewiring under removal of a node. (b)~A schematic diagram to calculate the probability that a randomly chosen out-link leads to a small out-component [square] with random bypass rewiring under removal of a node.}
\end{figure}

From now on, we formulate the equations corresponding to Eqs.~(\ref{eq:H_in_1_H_in_1}), (\ref{eq:H_out_1_H_out_1}), (\ref{eq:u_in}), and (\ref{eq:u_out}) with considering random bypass rewiring.
Based on the idea seen in Fig.~\ref{fig:gen_bypass_in_out} which illustrates random bypass rewiring on an infinite directed random network,
\begin{subequations}
\begin{flalign}
\qquad & H_{\text{in},1}(x) = \sum_{j=0}^{\infty} \sum_{k=0}^{\infty} \frac{ k p_{j,k} }  {\langle k \rangle}  \bigg\{ \phi_{j,k} [H_{\text{in},1}(x)]^j  + (1 - \phi_{j,k}) \bigg[ \min(\frac{j}{k},1) H_{\text{in},1}(x) + \max(0, \frac{k-j}{k}) \bigg] \bigg\} \text{,} \label{eq:H_in_1_bypass} & \\
\qquad & H_{\text{out},1}(x) = \sum_{j=0}^{\infty} \sum_{k=0}^{\infty} \frac{ j p_{j,k} }   {\langle j \rangle} \bigg\{ \phi_{j,k} [H_{\text{out},1}(x)]^k + (1 - \phi_{j,k}) \bigg[ \min(\frac{k}{j},1) H_{\text{out},1}(x) + \max(0, \frac{j-k}{j}) \bigg] \bigg\} \text{,} & \label{eq:H_out_1_bypass}
\end{flalign}
\end{subequations}
\begin{subequations}
\begin{flalign}
\qquad & u_\text{in} = H_{\text{in},1}(1) = f_3(u_\text{in}) = \sum_{j=0}^{\infty} \sum_{k=0}^{\infty} \frac{ k p_{j,k} }   {\langle k \rangle} \bigg\{ \phi_{j,k} u_\text{in}^j + (1 - \phi_{j,k}) \bigg[ \min(\frac{j}{k},1) u_\text{in} + \max(0, \frac{k-j}{k}) \bigg] \bigg\} \text{,} & \label{eq:u_in_bypass} \\
\qquad & u_\text{out} = H_{\text{out},1}(1) = f_4(u_\text{out}) = \sum_{j=0}^{\infty} \sum_{k=0}^{\infty} \frac{ j p_{j,k} }  {\langle j \rangle} \bigg\{ \phi_{j,k} u_\text{out}^k + (1 - \phi_{j,k}) \bigg[ \min(\frac{k}{j},1) u_\text{out} + \max(0, \frac{j-k}{j}) \bigg] \bigg\} \text{,} & \label{eq:u_out_bypass}
\end{flalign}
\end{subequations}
are derived.

In the case of undirected networks, the average probability that a randomly chosen link is not connected to the giant component is
\begin{flalign}
\qquad & u = \sum_{k=0}^{\infty}{q_k \phi_{k+1} u^k} + u \sum_{k=0}^{\infty}{q_k (1 - \phi_{k+1})} + (1 - u) \sum_{k=0}^{\infty} { \frac{p_{2k+1} (1-\phi_{2k+1})}{\sum_{k'=1}^{\infty}{k'p_{k'}}} } \text{,} & \label{eq:u_bypass}
\end{flalign}
for $q_k = (k+1)p_k / \sum_{k=0}^{\infty} {k p_k}$ \cite{park16}.

The self-consistent equations like Eqs.~(\ref{eq:u_in}), (\ref{eq:u_out}), (\ref{eq:u_in_bypass}), and (\ref{eq:u_out_bypass}) can be solved as follows by the fixed-point iteration \cite{burden}.
Iterating
\begin{subequations}
\begin{flalign}
\qquad & u_{\text{in},i+1} = f_1(u_{\text{in},i}) \text{,} & \label{eq:u_in_i_f1} \\
\qquad & u_{\text{out},i+1} = f_2(u_{\text{out},i}) \text{,} & \label{eq:u_out_i_f2}
\end{flalign}
\end{subequations}
\begin{subequations}
\begin{flalign}
\qquad & v_{\text{in},i+1} = f_3(v_{\text{in},i}) \text{,} & \label{eq:v_in_i_f3} \\
\qquad & v_{\text{out},i+1} = f_4(v_{\text{out},i}) \text{,} & \label{eq:v_out_i_f4}
\end{flalign}
\end{subequations}
for $u_{\text{in},0} = u_{\text{out},0} = v_{\text{in},0} = v_{\text{out},0} = 0$, $u_{\text{in},i}$, $u_{\text{out},i}$, $v_{\text{in},i}$, and $v_{\text{out},i}$ approaches to $\bar u_\text{in}$, $\bar u_\text{out}$, $\bar v_\text{in}$, and $\bar v_\text{out}$, respectively, as $i$ goes to infinity, for
\begin{subequations}
\begin{flalign}
\qquad & \bar u_\text{in} = f_1(\bar u_\text{in}) \text{,} & \label{eq:u_in_f1} \\
\qquad & \bar u_\text{out} = f_2(\bar u_\text{out}) \text{,} & \label{eq:u_out_f2}
\end{flalign}
\end{subequations}
\begin{subequations}
\begin{flalign}
\qquad & \bar v_\text{in} = f_3(\bar v_\text{in}) \text{,} & \label{eq:v_in_f3} \\
\qquad & \bar v_\text{out} = f_4(\bar v_\text{out}) \text{.} & \label{eq:v_out_f4}
\end{flalign}
\end{subequations}
Since
\begin{subequations}
\begin{flalign}
\qquad & f_1(u_\text{in}) \geq f_3(u_\text{in}) \text{,} & \label{eq:f_1_u_f_3_u} \\
\qquad & f_2(u_\text{out}) \geq f_4(u_\text{out}) \text{,} & \label{eq:f_2_u_f_4_u}
\end{flalign}
\end{subequations}
hold from $u_\text{in} \min(\frac{j}{k},1) + \max(0,\frac{k-j}{k}) \leq 1$ and $u_\text{out} \min(\frac{k}{j},1) + \max(0,\frac{j-k}{j}) \leq 1$ for $j, k \geq 0$ and $0 \leq u_\text{in}, u_\text{out} \leq 1$,
\begin{subequations}
\begin{flalign}
\qquad & u_{\text{in},i} \geq v_{\text{in},i} \text{,} & \label{eq:u_in_v_i} \\
\qquad & u_{\text{out},i} \geq v_{\text{out},i} \text{,} & \label{eq:u_out_v_i}
\end{flalign}
\end{subequations}
are derived for all $i$ from Eqs.~(\ref{eq:u_in_i_f1}), (\ref{eq:u_out_i_f2}), (\ref{eq:v_in_i_f3}), and (\ref{eq:v_out_i_f4}).
Therefore, $S_\text{s}$ on an infinite directed random network with random bypass rewiring is always equal to or larger than without random bypass rewiring; the percolation threshold on an infinite directed random network with random bypass rewiring is always equal to or smaller than without random bypass rewiring.

For even degree infinite undirected random networks; that is,
\begin{flalign}
\qquad & p_{2k+1} = 0 \text{,} & \label{eq:p_2k_1}
\end{flalign}
the probability that a randomly chosen node belongs to the giant component is
\begin{flalign}
\qquad & S = \sum_{k=0}^{\infty} {p_k \phi_k (1 - u^k) } = \sum_{k=0}^{\infty}{p_k \phi_k} = \phi \text{,} & \label{eq:S}
\end{flalign}
since Eq.~(\ref{eq:u_bypass}) is reduced to
\begin{flalign}
\qquad & u = \sum_{k=0}^{\infty}{q_k \phi_{k+1} u^k} + u \sum_{k=0}^{\infty}{q_k (1 - \phi_{k+1})} \text{,} & \label{eq:u_bypass_reduced}
\end{flalign}
and $u=0$ is the smallest non-negative real solutions of Eq.~(\ref{eq:u_bypass_reduced}) \cite{park16}.
Similarly, if every node, except a fixed number of nodes, on infinite directed random networks has in-degree equal to the out-degree; that is,
\begin{flalign}
\qquad & p_{j,k} = 0 \text{ for } j \neq k \text{,} & \label{eq:p_j_k_0}
\end{flalign}
holds,
Eqs.~(\ref{eq:u_in_bypass}) and (\ref{eq:u_out_bypass}) are reduced to
\begin{subequations}
\begin{flalign}
\qquad & u_\text{in} = \sum_{j=0}^{\infty} \sum_{k=0}^{\infty} \frac{ k p_{j,k} }   {\langle k \rangle} \bigg[ \phi_{j,k} u_\text{in}^j + (1 - \phi_{j,k}) u_\text{in} \bigg] \text{,} & \label{eq:u_in_bypass_reduced} \\
\qquad & u_\text{out} = \sum_{j=0}^{\infty} \sum_{k=0}^{\infty} \frac{ j p_{j,k} }  {\langle j \rangle} \bigg[ \phi_{j,k} u_\text{out}^k + (1 - \phi_{j,k}) u_\text{out} \bigg] \text{.} & \label{eq:u_out_bypass_reduced}
\end{flalign}
\end{subequations}
Therefore, the smallest non-negative real solutions of Eqs.~(\ref{eq:u_in_bypass_reduced}) and (\ref{eq:u_out_bypass_reduced}) are $u_\text{in} = u_\text{out} = 0$ for which Eq.~(\ref{eq:S_s}) corresponds to
\begin{flalign}
\qquad & S_\text{s} = \sum_{j=0}^{\infty} { \sum_{k=0}^{\infty} { p_{j,k} \phi_{j,k} }    } = \phi \text{,} & \label{eq:S_s_phi}
\end{flalign}
from Eq.~(\ref{eq:phi_p_j_k}).
Therefore, $S_\text{s}$ is equal to $\phi$, and the percolation threshold is $0$ on an infinite directed random network with random bypass rewiring for Eq.~(\ref{eq:p_j_k_0}); that is, infinite directed random networks are extremely robust with random bypass rewiring for arbitrary $\phi_{j,k}$ where Eq.~(\ref{eq:p_j_k_0}) holds.

\subsection{Simulation results}
To simulate attacks, a node with the highest product of in-degree and out-degree is firstly removed and nodes are removed one by one in decreasing order of the product of in-degree and out-degree while randomly chosen nodes are removed one by one in the case of failures ($\phi_{j,k} = \phi$).
For a numerical simulation of attacks for Eqs.~(\ref{eq:u_in}), (\ref{eq:u_out}), (\ref{eq:u_in_bypass}), and (\ref{eq:u_out_bypass}),
\begin{flalign}
\qquad & \phi_{j,k} = \begin{cases}
1 \text{,} & \sum_{\{ j'k' \leq jk \} } p_{j',k'} < \phi \\
\frac{\phi - \sum_{\{ j'k' \leq jk-1 \} } p_{j',k'} } {\sum_{\{ j'k' = jk \}} p_{j',k'}} \text{,} & \sum_{\{ j'k' \leq jk-1 \}} p_{j',k'} < \phi \text{ and} \sum_{\{ j'k' \leq jk \}} p_{j',k'} \geq \phi \\
0 \text{,} & \text{otherwise}
\end{cases} &
\end{flalign}
is set for given average occupation probability $\phi$.
In the simulations, in-degree and out-degree of each node is not recalculated while nodes are removed.
To simulate random bypass rewiring, each pair of in-links and out-links of the removed node are randomly chosen and rewired until there is no pair to match.

\begin{figure}[]
\center
\subfigure[]{
	\includegraphics[width=3.3in]{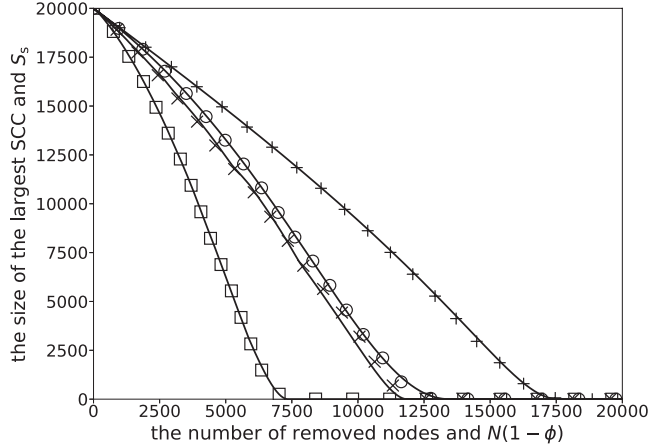}
	\label{fig:sim_num_2}
}
\subfigure[]{
	\includegraphics[width=3.3in]{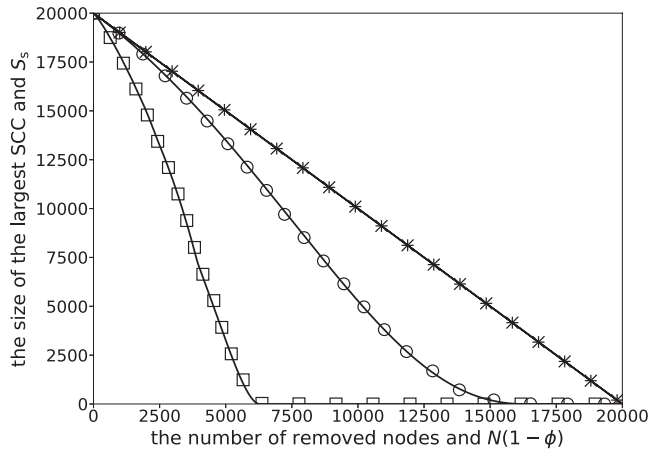}
	\label{fig:sim_num_even_2}
}
\caption{\label{fig:sim2} The size of the largest SCC (strongly connected component) with respect to the number of removed nodes under failures [circles (plus signs)] and attacks [squares (crosses)] without (with) random bypass rewiring. The solid lines are for numerically calculated $S_\text{s}$ with respect to $N(1- \phi)$ from Eqs.~(\ref{eq:S_s}), (\ref{eq:u_in}), (\ref{eq:u_out}), (\ref{eq:u_in_bypass}), and (\ref{eq:u_out_bypass}) on an infinite directed random network with the same degree distribution. (a)~On the directed network randomly generated by the configuration model with in-degree distribution $p_{\text{in},j} \sim j^{-3}$, out-degree distribution $p_{\text{out},k} \sim k^{-3}$, $N = 20000$ nodes, and $M = 61438$ links where $j$ and $k$ are almost uncorrelated. (b)~On the directed network randomly generated by the configuration model with in-degree distribution $p'_{\text{in},j} = p_{\text{in},j}$, out-degree distribution $p'_{\text{out},k} = p_{\text{out},k}$, $N = 20000$ nodes, and $M=61438$ links where each node has in-degree equal to the out-degree ($p'_{j,k}=0$ for $j \neq k$). Two straight lines for random bypass rewiring are overlapped.}
\end{figure}

The directed network for Fig.~\ref{fig:sim_num_even_2} is randomly generated by in-degree distribution $p'_{\text{in},j} = p_{\text{in},j}$, out-degree $p'_{\text{out},k} = p_{\text{out},k}$, and degree distribution $p'_{j,k}=0$ for $j \neq k$ where $p_{\text{in},j}$ and $p_{\text{out},k}$ are the in-degree distribution and the out-degree distribution of the original directed network for Fig.~\ref{fig:sim_num_2}, respectively.
In other words, the directed network for Fig.~\ref{fig:sim_num_even_2} has a perfect positive correlation between in-degree and out-degree while the original directed network for Fig.~\ref{fig:sim_num_2} has an almost uncorrelated relationship between in-degree and out-degree.

Figure~\ref{fig:sim_num_even_2} shows that almost all nodes except the removed nodes on the directed network are strongly connected by random bypass rewiring.
The percolation threshold with random bypass rewiring in Fig.~\ref{fig:sim_num_even_2} is close to $0$, while the percolation threshold in Fig.~\ref{fig:sim_num_2} is not.

Without random bypass rewiring, the original directed network for Fig.~\ref{fig:sim_num_2} is more robust than the directed network for Fig.~\ref{fig:sim_num_even_2} under attacks while it is not under failures, even though the directed network for Fig.~\ref{fig:sim_num_even_2} is randomly generated by the same in-degree distribution and out-degree distribution of the original directed network for Fig.~\ref{fig:sim_num_2}.
On the other hand, with random bypass rewiring, the directed network for Fig.~\ref{fig:sim_num_even_2} is more robust than the original directed network for Fig.~\ref{fig:sim_num_2} against removal of nodes including failures and attacks.
In other words, the directed network for Fig.~\ref{fig:sim_num_even_2} is more robust than the original directed network for Fig.~\ref{fig:sim_num_2} with random bypass rewiring, while the original directed networks for Fig.~\ref{fig:sim_num_2} is more robust than the directed network for Figure~\ref{fig:sim_num_even_2} under attacks without random bypass rewiring.

\section{Discussion} \label{sec:discussion}
\subsection{Generalized results on infinite random networks}
From Eq.~(\ref{eq:S}) [Eq.~(\ref{eq:S_s_phi})], $S$ [$S_\text{s}$] is equal to $\phi$ for Eq.~(\ref{eq:p_2k_1}) [Eq.~(\ref{eq:p_j_k_0})] on an infinite undirected [directed] random network.
$S$ [$S_\text{s}$] is always smaller than $\phi$ with random bypass rewiring if Eq.~(\ref{eq:p_2k_1}) [Eq.~(\ref{eq:p_j_k_0})] does not hold on an infinite undirected [directed] random network.
Therefore, $S$ [$S_\text{s}$] is equal to $\phi$ with random bypass rewiring if and only if Eq.~(\ref{eq:p_2k_1}) [Eq.~(\ref{eq:p_j_k_0})] holds on an infinite undirected [directed] random network; that is, an infinite undirected [directed] random network is extremely robust with random bypass rewiring if and only if Eq.~(\ref{eq:p_2k_1}) [Eq.~(\ref{eq:p_j_k_0})] holds.

\subsection{Generalized results on finite networks} \label{sec:finite}

From Eqs.~(\ref{eq:p_2k_1}) and (\ref{eq:p_j_k_0}), we recall a necessary condition for the existence of Eulerian cycles on a finite network \cite{bollobas}.
In the case of finite networks, Eq.~(\ref{eq:p_2k_1}) [Eq.~(\ref{eq:p_j_k_0})] is interpreted that an undirected [directed] network has no node with odd degree [different in-degree and out-degree].

From the interpretation of Eqs.~(\ref{eq:p_2k_1}) and (\ref{eq:p_j_k_0}), we hypothesized that a finite network has a strongly connected spanning subnetwork which has an Eulerian path or cycle if and only if there exists an way of bypass rewiring to strongly connect all nodes except the removed nodes on the finite network for every combination of removed nodes.

If there exists a set of bypass-rewired links which strongly connect all nodes except the removed nodes on a finite network for every combination of removed nodes, the finite network has a strongly connected spanning subnetwork which is composed of the bypass-rewired links and all nodes on the network.
Since each bypass-rewired link (in-link) of a node on the strongly connected spanning subnetwork is paired with another bypass-rewired link (a bypass-rewired out-link) of the node, the number of the bypass-rewired links (in-links) of every node is even (is equal to the number of the bypass-rewired out-links); that is, the strongly connected spanning subnetwork with the bypass-rewired links has an Eulerian path or cycle.

Since the degree (in-degree) of every node, except the starting and ending node of the Eulerian path, on a strongly connected spanning subnetwork which has an Eulerian path or cycle is even (equal to the out-degree), each link (in-link) of a node on the Eulerian path or cycle can be bypass-rewired to the next sequenced link (out-link) [a link (an in-link) of the next sequenced node] of the Eulerian path or cycle which has sequentially ordered links; that is, there exists at least one way of bypass rewiring to connect all nodes except the removed nodes on the finite network for every combination of removed nodes.
For example, the $i$-th link of an Eulerian path or cycle with $L$ links can be bypass-rewired to the $(i+1)$-th link of the Eulerian path or cycle for $1 \leq i < L$ (and the $L$-th link of the Eulerian cycle can be bypass-rewired to the $1$st link of the Eulerian cycle since the starting and ending node of an Eulerian cycle are same), as seen in Fig.~\ref{fig:compuserve}.

In consequence, it is derived that a finite network has a strongly connected spanning subnetwork which has an Eulerian path or cycle if and only if there exists an way of bypass rewiring to make the finite network extremely robust for every combination of removed nodes; it is naturally satisfied that Eulerian networks are extremely robust for every combination of removed nodes.

Even though an way of bypass rewiring based on the sequentially ordered links of a strongly connected spanning subnetwork which has an Eulerian path or cycle is enough to strongly connect all nodes on the network, not yet rewired links of each node can be additionally bypass-rewired to improve connectivity.
For example, link F-B and link B-I of node B in Fig.~\ref{fig:compuserve} can be additionally bypass-rewired to improve connectivity.

\subsection{A real-world example: the Internet topology}

\begin{figure}
\center
\includegraphics[width=3in]{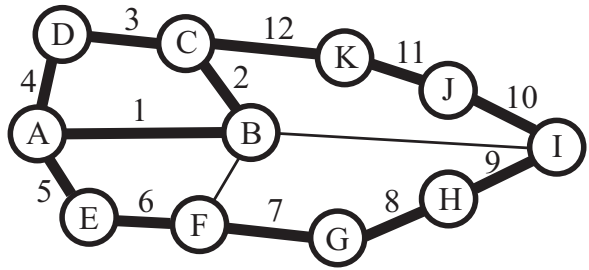}
\caption{\label{fig:compuserve} Compuserve topology, a part of the Internet topology, in U.S. on Jan 2011, publicly available in \cite{topologyzoo}. The spanning subnetwork with the bold links has an Eulerian path with $L=12$ links. Each number of the bold links denotes the sequential order of the Eulerian path.}
\end{figure}

From the consequence derived in Section \ref{sec:finite}, the network in Fig.~\ref{fig:compuserve} is extremely robust for every combination of removed nodes since the network has a spanning subnetwork which has an Eulerian path (A-B-C-D-E-F-G-H-I-J-K-C).
To connect all nodes except the broken nodes on the network in Fig.~\ref{fig:compuserve} for every combination of removed nodes, the bypass rewiring policy is defined as follows, based on the sequentially ordered links of the Eulerian path of the spanning subnetwork.
When node B is broken, the $1$st link (link A-B) should be bypass-rewired to the $2$nd link (link B-C) and link F-B can be additionally bypass rewired to link B-I.
When node C is broken, the $2$nd link (link B-C) should be bypass-rewired to the $3$rd link (link C-D).
When node D is broken, the $3$rd link (link C-D) should be bypass-rewired to the $4$th link (link D-A).
When node A is broken, the $4$th link (link D-A) should be bypass-rewired to the $5$th link (link A-E).
When node E is broken, the $5$th link (link A-E) should be bypass-rewired to the $6$th link (link E-F).
When node F is broken, the $6$th link (link E-F) should be bypass-rewired to the $7$th link (link F-G).
When node G is broken, the $7$th link (link F-G) should be bypass-rewired to the $8$th link (link G-H).
When node H is broken, the $8$th link (link G-H) should be bypass-rewired to the $9$th link (link H-I).
When node I is broken, the $9$th link (link H-I) should be bypass-rewired to the $10$th link (link I-J).
When node J is broken, the $10$th link (link I-J) should be bypass-rewired to the $11$th link (link J-K).
When node K is broken, the $11$th link (link J-K) should be bypass-rewired to $12$th link (link K-C).

The network in Fig.~\ref{fig:compuserve} can be fragmented into two parts, a part (node A, B, C, D, E, and F) and another part (node H, I, and J), when node G and K are broken under failures or attacks or is under repair.
According to the bypass rewiring policy defined above, link F-G would be bypass-rewired to link G-H, link J-K would be bypass-rewired to link K-C, and then all nodes except the broken node (node G and K) on the network can be connected.

\section{Conclusions} \label{sec:conclusion}
In summary, we have introduced a concept of bypass rewiring on directed networks and conducted analytical and numerical investigations of random bypass rewiring with double generating function formalisms and simulations.
The results have shown that random bypass rewiring improves robustness of infinite directed (undirected) random networks under removal of nodes including failures and attacks.
In particular, random bypass rewiring guarantees extreme robustness of infinite directed (undirected) random networks for arbitrary occupation probabilities where in-degree of every node except a fixed number of nodes is equal to the out-degree (every node except a finite number of nodes has even degree); bypass rewiring makes the percolation threshold $0$.
From the above results, we have derived that a finite network has a strongly connected spanning subnetwork which has an Eulerian path or cycle if and only if there exists an way of bypass rewiring to guarantee extreme robustness of the finite network for every combination of removed nodes; Eulerian networks are extremely robust for every combination of removed nodes.

If the Internet topology has a spanning subnetwork which has an Eulerian path or cycle like Compuserve topology in Fig.~\ref{fig:compuserve}, the Internet can be extremely robust against breakdown of routers (nodes) since links of a broken router on the Internet can be bypass-rewired by an engineer or equipment.
Therefore, it is suggested that bypass rewiring equipment and routers be implemented and deployed on the Internet.
In addition, variations on bypass rewiring theory including optimal bypass rewiring algorithms and more applications to many fields like electric circuits and power grids are expected.

\section*{References}



\bibliographystyle{elsarticle-num} 
\bibliography{paper_preprint}






\end{document}